\documentclass[10pt]{llncs}

\usepackage{amsmath}
\usepackage{semantic}
\usepackage{amssymb}
\usepackage{stmaryrd}

\usepackage{mathtools}
\usepackage{comment}

\usepackage[linesnumbered]{algorithm2e}

\usepackage{tikz}
\usetikzlibrary{arrows,automata}
\usetikzlibrary{decorations.pathmorphing}

\usepackage{multicol}
\newcommand{\Stop}{\mathrm{\mathbf{stop}}}
\newcommand{\Def}{\mathrm{\mathbf{def~}}}
\newcommand{\seq}[1]{>\hspace{-0.9mm}#1\hspace{-0.9mm}>}
\newcommand{\prune}[1]{<\hspace{-0.9mm}#1\hspace{-0.9mm}<}

\newcommand{\lin}[1]{Lin(#1)}

\newcommand{\bibar}[1]{\overline{\overline{#1}}}

\newcommand{\vwhite}{\phantom{X}}

\newcommand{\ind}{~~~}%\phantom{aaa}}

\begin{document}

\title{Tracking Causal Dependencies in Web Services Orchestrations
  Defined in ORC}
\author{Matthieu Perrin \and Claude Jard \and Achour Most\'efaoui}
\institute{Universit\'e de Nantes, LINA, 2 rue de la Houssini\`{e}re,
  F-44322 Nantes, France\\
  ``first name''.``last name''@univ-nantes.fr
\footnote{This work has been partially supported by a French government support
granted to the CominLabs excellence laboratory (Project 
\emph{DeSceNt:Plug-based Decentralized Social Network}) 
and managed by the French
National Agency for Research (ANR) in the "Investing for the Future"
program under reference Nb. ANR-10-LABX-07-01. It was also 
partially funded by the French ANR project SocioPlug
(ANR-13-INFR-0003).}}

\maketitle
\begin{abstract}
This article shows how the operational semantics of a language like
ORC can be instrumented so that the execution of a program produces
information on the causal dependencies between events. The concurrent
semantics we obtain is based on asymmetric labeled event structures. The
approach is illustrated using a Web service orchestration instance
and the detection of race conditions.
\end{abstract}

\vspace{-0.6cm}
\section{Introduction}
\label{section:introduction}

Several languages have been proposed to program applications based on
Web service orchestrations (BPEL \cite{andrews2003business} is
probably one of the best known). The present work is based on Orc
\cite{orc,Kitchin2009}, an orchestration language whose definition
is based on a mathematical semantics, which is needed to define
precisely the notion of causality. Orc is designed over the notion of \emph{sites},
a generalization of functions that can encapsulate any kind of externally defined 
web sites or services as well as Orc expressions.
As usual for languages, the operational
semantics of Orc was defined as a labeled transition
system. Such semantics produces naturally sets of sequential traces,
which explicitly represent the observable behaviors
of an Orc program~\cite{KCM06}. 

Finding the causal dependencies in a program is very useful for
error detection. In a non-deterministic concurrent context, this analysis
cannot be based solely on the static structure of the program and
requires execution. Dependencies are also very difficult to extract
from a sequential record without additional information to unravel the
interleaving of events. This is especially true for the analysis of
QoS or of non functional properties, like timing constraints
derived from the critical path of dependencies~\cite{ISOLA2006}.  
We consider any Orc program, which has been already parsed and expanded into
its Orc calculus intermediate form. In this program, we distinguish the actions,
which are the site calls, and the publications (return values of expressions). An event is the
occurrence of an action during the execution of the Orc program. The events
are linked by causal dependencies, that force the events to be executed in a
certain order. We can distinguish three kinds of dependencies:
\begin{itemize}
\item  the dependencies that are imposed by the control 
flow of the program defined
by the semantics of the Orc combinators and imposed by the binding mechanism of Orc variables;
\item  the dependencies that are provided by the server executing the
  site calls. These external dependencies are not part of the Orc description, but
could be returned by the site. We will
consider at least that the possible return of a site call is directly caused by this
call;
\item the dependencies induced by preemption (the pruning operator of Orc).
\end{itemize}
The method used in this article is to extend the standard structural
operational semantics (SOS~\cite{Plotkin2004})
to rewrite extended expressions, in which additional
information has been added to compute causal and weakly-causal dependencies.
This information is also made visible by extending
the labeling of transitions. Concurrency is just the complement of
the weak-causal relation, and conflicts are defined by cycles in this relation.
Capturing causality and concurrency by instrumenting the semantics rules is a difficult job. This is mainly due to the fact that these relationships are global and therefore difficult to locate on the syntactic forms. 
The solution is to keep information about the causal past in a context associated with each rule. 
We build the necessary links between different contexts during the execution of rules.
The aim is that such instrumented semantics
reproduces the standard behavior of the program while
calculating the additional information needed to track concurrency,
causality and conflicts between the events produced by the execution.

After this introduction, the article presents the contribution compared to
the existing works. Section \ref{section:orc} presents the Orc language from the
perspective of its core calculus and its operational semantics,
illustrated using an example of orchestration of Web services. Section \ref{section:instrumentation}
presents our proposed instrumented semantics based on the construction
of event structures, giving the concurrent semantics of Orc. This
section sets out the formal correctness of the approach stating that
this new semantics produces the same executions as the standard
semantics. Before concluding the paper, Section \ref{section:application} reuses the example of Section \ref{section:orc}
to show the causal structure obtained from its execution in the instrumented
semantics and how this can be used to find errors.

\section{Related work}
\label{section:related}

The need to dynamically trace the causal dependencies during the execution
of the a program in order to monitor, detect errors or analyze
performances is well recognized for concurrent applications. Causality, seen as a
partial order~\cite{Lamport78}, can be tracked in different ways.
Some works are based on an instrumentation of either the underlying operating
system or the source code. For example, vector clocks have been widely
used by the distributed algorithms community in the context
of message-passing systems \cite{Fidge88}. The context
of Web service orchestrations is more complex as
a language like Orc can generate unbounded concurrency patterns. To our knowledge, the only
instrumentation made on programs is \cite{Rocsu2007}, based on Java byte-code. 
However, in the considered model, the only source of causality comes from
variable accesses.
The second approach is to change the semantics so that it produces causal
information which leads to a {\em concurrent semantics}. The challenge is then to
maintain a good form of equivalence with the original
semantics. Several debugging techniques rely on this principle,
especially for performing replay (\cite{Fase2014} is a good example
for a fragment of the Oz language).  The
most successful works in concurrent semantics were conducted on process algebra (e.g. pi-calculus
\cite{Boreale98}). Our contribution is in the same vein, but for the
Orc language, in the complex context of wide-area computing.
Other attempts of concurrent semantics for Orc based on event structures 
have already been published \cite{Rosario2007,Bruni2006}. They use  an ad hoc connection 
of Petri net diagrams or Join calculus. It is not clear how this
semantics can be implemented in practice at compile-time that transforms the source code into a concurrent model. 
An instrumented semantics solves this problem and allows to catch
causal dependencies at runtime. 

\section{The Orc Programming Language}
\label{section:orc}

\subsection{Core calculus}

Orc is a full programming language, that looks like a functional language
with many non-functional aspects to handle concurrency.
The interested reader can refer to \cite{wide-orc} concerning the ability of Orc
to design large-scale distributed applications.
The Orc programming language is designed over a process calculus: 
the Orc core calculus. 
All the conveniences offered in the full Orc language are derived from very few central 
concepts present in the calculus: sites and operators.
Values such as booleans, numbers and strings, arithmetic and logic operators,
as well as complex data types such as shared registers, are just
external sites.
Even choices are implemented through the use of sites \verb+ift+ and \verb+iff+, 
that publish a signal if their argument is true or false respectively. 
Besides sites, four operators are provided by the calculus to
orchestrate the execution. These operators describe the sequencing of
actions (``$f \seq{x} g$''), the launching of parallel threads (``$f
| g$''),  an original preemption operator (``pruning: $f \prune{x}
g$'') and an alternative in case of no response (``otherwise:  $f;g$'').
The full syntax of the calculus is specified by the grammar given in Figure \ref{figure:syntaxe}.
From now on, we denote by $Orc_s$ the set of the expressions 
allowed by this syntax. 
The expressions of the calculus that correspond to real Orc programs,
denoted by the set $Orc$,
are those that do not contain $?k$ and $\bot$ expressions.

\begin{figure}
$$
\begin{array}{lll}
  f, g, h\in \text{Expression}       & ::= &   p \| p(p) \| ?k \| f|g \| f \seq{x} g \| f \prune{x} g \| f;g \| D \# f \| \bot\\
  D\in \text{Definition}             & ::= &   \Def y(x) = f\\
  v\in \text{Orc~Value}              & ::= &   V \| D\\
  p\in \text{Parameter}              & ::= &   v \| \Stop \| x\\
  w\in \text{Response}               & ::= &   NT(v) \| T(v) \| Neg\\
  n\in \text{Hidden Label}           & ::= &   ?V_k(v) \| ?D \| h(\omega) \| h(!v)\\
  l\in \text{Label}                  & ::= &   !v \| n \| \omega
\end{array}
$$
\caption{The syntax of the Orc core calculus.}
\label{figure:syntaxe}
\end{figure}

There are two kinds of sites in Orc: the external ones, denoted $V$ in
the syntax, and the internal ones defined as an Orc expression with the syntax
$\Def y(x) = f \# g$ where $f$ is the body of the site and $g$
is the remaining of the program in which $y$ can be used as any site. 
For the sake of clarity, we consider in this work that the sites are curryfied, so they have exactly one argument. 
Site definitions are recursive,
which allows the same expressivity as any functional language. 
Calls to external sites
are strict, i.e. their arguments have to be bound before the site can be called, while an
internal site can be called immediately, and its arguments are evaluated lazily.
When an external site is called, it sends its responses to a placeholder $?k$.
A response can be either a non-terminating value $NT(v)$ if further responses are expected, or a 
terminating value $T(v)$ if this is the last publication of the site, or $Neg$ if the site terminates
without publishing any value. 
In $f|g$, the parallel composition expresses pure concurrency; 
$f$ and $g$ are run in parallel, their events are interleaved and 
the expression stops when both $f$ and $g$ have terminated.
Sequentiality can be expressed by the sequential operator, like in $f \seq{x} g$, where
the variable $x$ can be used in $g$.
Here, $f$ is started first, and then a new instance of $g[v/x]$, where $x$ is bound to $v$, 
is launched as a consequence of each publication of $v$. In $f\prune{x} g$, the 
pruning operator is used to express preemption. The variable $x$ can be used in $f$.
Both $f$ and $g$ are started, but $f$ is paused when it needs 
to evaluate $x$. When $g$ publishes a value, it is bound to $x$ in $f$, and $g$ 
is stopped. Other events that could have been produced by $g$ are preempted 
by the publication. For example, if $g$ is supposed to publish two values $a$ and $b$,
only one will be selected and published in each execution. We say that these two events are 
in conflict. The pruning operator is left-associative: in $f\prune{x}g\prune{y}h$, $f$, $g$ 
and $h$ are started in parallel, the first publication of $g$ is bound to $x$ and the first 
publication of $h$ is bound to $y$. 
The otherwise operator is used in $f;g$. In this expression, $f$ is first 
started alone and $g$ is started if and only if $f$ stops without publishing any value.
Finally, the $\Stop$ symbol can be used by the programmer exactly like a site or a variable to 
denote a terminated program. $\Stop$ still produces an event $\omega$ to notify
its parent expression that it has terminated. It then evolves into $\bot$, the inert final expression.
$?k$ and $\bot$ cannot be used directly.

\begin{figure*}[h]
\begin{center}
\begin{tabular}{lr}
\inference[(\textsc{Publish})]
          {\vwhite}
          {v\xrightarrow{!v} \Stop}
          [$v$ closed]
\\
\inference[(\textsc{Stop})]
          {\vwhite}
          {\Stop \xrightarrow{\omega} \bot}
&
\inference[(\textsc{StopCall})]
          {\vwhite}
          {\Stop(p)\xrightarrow{\omega} \bot}
\\
\inference[(\textsc{ExtStop})]
          {\vwhite}
          {V(\Stop)\xrightarrow{\omega} \bot}
&
\inference[(\textsc{ExtCall})]
          {\vwhite}
          {V(v)\xrightarrow{?V_k(v)} ?k}
          [$k$ fresh]
\\
\multicolumn{2}{l}{
\inference[(\textsc{DefDeclare})]
          {[D/y]f \xrightarrow{l} f'}
          {D \# f \xrightarrow{l} f'}
          [$D$ is \textbf{def} $y(x) = g$]
}
\\
\multicolumn{2}{l}{
\inference[(\textsc{IntCall})]
          {\vwhite}
          {D(p) \xrightarrow{?D} [D/y][p/x]g}
          [$D$ is \textbf{def} $y(x) = g$]
}
\\
\inference[(\textsc{ParLeft})]
          {f\xrightarrow{l} f'}
          {f|g\xrightarrow{l} f'|g}
          [$l\neq\omega$]
&
\inference[(\textsc{ResT})]
          {?k \text{~receives~} T(v)}
          {?k\xrightarrow{!v} \Stop}
\\
\inference[(\textsc{ParRight})]
          {g\xrightarrow{l} g'}
          {f|g\xrightarrow{l} f|g'}
          [$l\neq\omega$]
&
\inference[(\textsc{ResNt})]
          {?k \text{~receives~} NT(v)}
          {?k\xrightarrow{!v} ?k}
\\
\inference[(\textsc{ParStop})]
          {f\xrightarrow{\omega} \bot & g\xrightarrow{\omega} \bot}
          {f|g \xrightarrow{\omega} \bot}
&
\inference[(\textsc{ResNeg})]
          {?k \text{~receives~} Neg}
          {?k\xrightarrow{\omega} \bot}
\\
\inference[(\textsc{OtherV})]
          {f\xrightarrow{!v} f'}
          {f;g\xrightarrow{!v} f'}
&
\inference[(\textsc{SeqV})]
          {f \xrightarrow{!v} f'}
          {f\seq{x}g \xrightarrow{h(!v)} (f'\seq{x}g) | [v/x]g}
\\
\inference[(\textsc{OtherN})]
          {f\xrightarrow{n} f'}
          {f;g\xrightarrow{n} f';g}
&
\inference[(\textsc{SeqN})]
          {f\xrightarrow{n} f'}
          {f\seq{x}g \xrightarrow{n} f'\seq{x}g}
\\
\inference[(\textsc{OtherStop})]
          {f\xrightarrow{\omega} \bot}
          {f; g \xrightarrow{h(\omega)} g}
&
\inference[(\textsc{SeqStop})]
          {f\xrightarrow{\omega} \bot}
          {f \seq{x} g \xrightarrow{\omega} \bot}
\\
\inference[(\textsc{PruneV})]
          {g \xrightarrow{!v} g'}
          {f\prune{x}g \xrightarrow{h(!v)} [v/x]f}
&
\inference[(\textsc{PruneLeft})]
          {f\xrightarrow{l} f'}
          {f\prune{x}g \xrightarrow{l} f'\prune{x}g}
          [$l\neq\omega$]
\\
\inference[(\textsc{PruneN})]
          {g \xrightarrow{n} g'}
          {f\prune{x}g \xrightarrow{n} f\prune{x}g'}
&
\inference[(\textsc{PruneStop})]
          {g \xrightarrow{\omega} \bot}
          {f \prune{x} g \xrightarrow{h(\omega)} [\Stop/x]f}
\\
\end{tabular}
\end{center}
\caption{The rules of the operational semantics.}
\label{figure:operational_semantics}
\end{figure*}

\subsection{Illustration}

\begin{figure}
\begin{algorithm}[H]
$\Def$ find\_best(agencies, destination) = \\
\ind$\Def$ find\_offers() = \\
\ind\ind each(agencies) $>$ agency $>$ agency(destination) $>$ offer $>$ \\
\ind\ind (offers.add((offer, agency)) $|$ \\
\ind\ind (best\_offer.read() $>o>$ compare(o, offer) $>b>$\\
\ind\ind ift(b) $>x>$ (best\_agency.write(agency) $|$ best\_offer.write(offer)))) \hfill\#\\
\ind$\Def$ extend\_best() = \\
\ind\ind best\_agency.read() $>$ ba $>$ best\_offer.read() $>$ bo $>$ ba.exists(bo) $>$ b $>$\\
\ind\ind (ift(b) $>x>$ ba.get\_info(bo) $|$ iff(b) $>x>$ alarm("inconsistent"))\hfill\# \\
\ind$\Def$ sort\_offers(offers, best\_offer) = \\
\ind\ind offers.sort(); best\_offer.read() = offers.first() $>$b$>$ \\
\ind\ind (ift(b) $>x>$ offers $|$ iff(b) $>x>$ alarm("not best"))\hfill\#\\
\ind ((t $<$t$<$ (find\_offers() $|$ timer(2000))) $>t>$ \\
\ind ((e\_b, s\_o) $<$e\_b$<$ extend\_best() $<$s\_o$<$ sort\_offers()))\\
\ind $<$offers$<$ Stack()\\
\ind $<$best\_offer$<$ (Register() $>$r$>$ r.write(null); r)\\
\ind $<$best\_agency$<$ Register()\hfill\#
\end{algorithm}
\caption{Identification of the best offers for a destination proposed by a pool of agencies.}
\label{figure:illustration_orc}
\end{figure}

We now illustrate the use of Orc in Figure \ref{figure:illustration_orc}. 
This program defines the internal site \verb+find_best(agencies, destination)+ that computes the best
offers proposed by the agencies listed in \verb+agencies+ for the
destination given as a parameter.
It publishes a unique value that is a pair composed of the best offer augmented
with additional information and the list of other offers sorted by price.
The program is composed of three internal sites.
It uses three shared objects, that are created in lines 15 to 17: the stack \verb+offers+
and the registers \verb+best_offer+ and \verb+best_agency+. At line~16, a new register is created through
a call to the site \verb+Register()+ and is bound to the variable \verb+r+. It is then initialized to a default value:
\verb+r.write(null)+ that can be seen as a shortcut for \verb+r("write")>w>w(null)+, so the shared register is a site 
that can publish its accessors when it is called. As writing in a register does not publish any value, 
the otherwise operator is finally used to bound the value to \verb+best_offer+. 
At line 11, the site \verb+find_offers+ can be started before the
variables are created (left hand side of pruning operators). \verb+each+ 
publishes in parallel all the sites contained into the stack \verb+agencies+, so 
all known agencies have to publish their offers. Each time a new offer is found, it is added into 
\verb+offer+ and its price is compared to the current best known offer.
The test is first evaluated and passed as an argument
to \verb+ift+. If  true, the program publishes a signal and the registers can be updated.
\verb+find_offers+ does not publish any value. In parallel with its call, we start a timer that 
publishes after 2 seconds a signal. The signal will halt this part of the program thanks to the pruning operator, and starts the line 14, thanks to the sequential operator. 
Line~14 calls both \verb+extend_best+ and \verb+sort_offers+ and publishes the result
when both sites have published. The two sites call an external site
either to sort the offers or to get extra information about the best offer,
and they perform a test that raises an alarm if something wrong is detected.

\begin{figure}[t]
\begin{multicols}{3}
\begin{enumerate}
\item each([A1, A2])
  \item timer(2000)
  \item new\_register()
  \item new\_register()
  \item A1(D)
  \item r.write(null)
  \item best\_offer.read()
  \item new\_stack()
  \item offers.add(O1)
  \item A2(D)
  \item offers.add(O2)
  \item compare(null, 01)
  \item best\_offer.read()
  \item compare(null, 02)
  \item ift(true)
  \item ift(true)
  \item best\_offer.write(O2)
  \item best\_offer.write(O1)
  \item best\_agency.write(A1)
  \item best\_agency.write(A2)
  \item best\_agency.read()
  \item best\_offer.read()
  \item A2.exists(O1)
  \item iff(false)
  \item ift(false)
  \item alarm(``inconsistent'')
  \item offers.sort()
  \item best\_offer.read()
  \item offers.first()
  \item =(O1, O2)
  \item iff(false)
  \item ift(false)
  \item alarm(``not best'')
\end{enumerate}
\end{multicols}
\caption{A possible execution for the program in Figure \ref{figure:illustration_orc} where agencies = $[A1, A2]$ 
and destination = $D$. Each agency publishes an offer $O1$ and $O2$ respectively. For the sake of space and clarity, 
we only show site calls in this execution.}
\label{figure:sequential_execution}
\end{figure}

Figure \ref{figure:sequential_execution} shows a possible trace of the
program of Figure \ref{figure:illustration_orc}. 
In this example, both alarms are due to inconsistencies in the shared registers.
To avoid the alarm ``inconsistent'', it is necessary to write into 
\verb+best_offer+ and \verb+best_agency+ atomically, and to avoid the other alarm, 
the comparison with the current value of \verb+best_offer+ and its edition should be
atomic. The event \verb+best_offer.write(01)+ is a cause for both alarms, but it
is impossible to detect it in the sequential trace without any information about causality.

\section{Instrumented Semantics}
\label{section:instrumentation}

\subsection{Method}

SOS specifications take the form of a set of inference rules that
define the valid transitions of a composite piece of syntax in terms
of the transitions of its components. Rewriting transforms terms by
executing a rule (it may be a non-deterministic transition in case of
multiple alternatives). The successive transitions represent the program
behavior. This may produce a sequence of values, that can be brought by
the labeling of rules.
Our approach is based on an
instrumentation of the rules, that appends additional information to the
labels in order to track the partial order of events. 
Actually, a label in the instrumented semantics is a tuple $e = (e_k, e_l, e_c, e_a)$, where
$e_k$ is an identifier taken in a countable set $K$, that is unique for the execution,
$e_l$ is a label similar to those of the standard semantics and $e_c$ and $e_a$
contain the finite sets of the identifiers of the causes and the weak causes of the event, 
respectively. Informally, an event $e$ is a cause of $e'$ if $e$ always
happens before $e'$, regardless of the scheduling chosen by the system. 
Similarly, $e$ is a weak cause of $e'$ if $e'$ can never happen after $e$, 
either because $e$ is one of its causes or because $e'$ preempts $e$.

In order to record the information concerning the past of an expression, 
we enrich the language with a new syntactic construction: $\langle f, c, a \rangle_L$ means 
that $c$ and $a$ are the causes of the Orc instrumented expression $f$.
Thus, if $f$ has $c$ and $a$ as causes and if it can evolve into $f'$, this transition
should also have $c$ and $a$ as causes. The index $L$ expresses the kind of events
that can activate the rule: $!v$ matches any publication,
$l$ stands for any label and $\omega$ means that $c$ and $a$ are only the causes
of the termination of the program. 
We also consider that the external sites track causality themselves, as
an internally-defined function would do. It makes sense as some sites (e.g. $+$)
handle their calls independently, while others (e.g. shared registers, 
management library) induce more complex causality patterns between the calls. 
Hence, the responses we get include this additional information. The verification of 
these responses is not the subject here, and we suppose them to be correct by hypothesis.

Apart from the introduction of the instrumentation construction and the new information in the responses,
the syntax of the instrumented expressions (Figure \ref{fig:instr_syntax}) is very similar to the regular one.
The set of all the expressions allowed by this extended syntax is $Orc_i$.
We can notice that every valid Orc program is also a valid
instrumented expression, which means that the instrumented 
semantics can be applied without program transformation. 

\begin{figure}[t]
  \begin{displaymath}
    \begin{array}{lll}
      f, g, h\in \text{Expression}       & ::= &   p \| p(p) \| ?k \| f|g \\
                                         &     &   \| f \seq{x} g \| f \prune{x} g \| f;g \\
                                         &     &   \| D \# f \| \bot \| \langle f, K, K \rangle_L \\
      D\in \text{Definition}             & ::= &   \Def y(x) = f\\
      v\in \text{Orc~Value}              & ::= &   V \| D\\
      p\in \text{Parameter}              & ::= &   v \| \Stop \| x \| \langle p, K, K \rangle_L \\
      w\in \text{Response}               & ::= &   NT(v, K, K) \| T(v, K, K) \\
                                         &     &   \| Neg(K, K)\\
      n\in \text{Hidden Label}           & ::= &   ?V_k(v) \| ?D \| h(\omega) \| h(!v)\\
      l\in \text{Label}                  & ::= &   !v \| n \| \omega
    \end{array}
  \end{displaymath}
  \caption{The extended syntax of the instrumented semantics.}
  \label{fig:instr_syntax}
\end{figure}

\subsection{Labeled Asymmetric Event Structure}
\label{subsection:LAES}

Labeled asymmetric event structures (LAES) \cite{winskel} are natural objects to 
represent concurrent executions in a compact way. 

\begin{definition}[Labelled asymmetric event structure]
  A \emph{labelled asymmetric event structure} (LAES) is a tuple $(E, L, \le, \nearrow, \Lambda)$. 
  \begin{itemize}
  \item $E$ is a set of \emph{events},
  \item $L$ is a set of \emph{labels},
  \item $\le$, \emph{causality} is a partial order on $E$,
  \item $\nearrow $, \emph{weak causality} is a binary relation on $E$,
  \item $\Lambda : E \mapsto L $ is the \emph{labelling function},
  \item each $e\in E$ has a finite \emph{causal history} $[e] = \{e'\in E | e'\le e\}$,
  \item for all events $e < e'\in E, e \nearrow e'$, where $<$ is the 
    irreflexive restriction of $\le$,
  \item for all $e\in E$, $\nearrow\cap[e] \times [e]$, the restriction of weak causality 
    to the causal history of $e$, is acyclic.
  \end{itemize}
\end{definition}

We also define an induced \emph{conflict} relation $\#_a$ as the smallest set of finite parts of $E$ such that:
for $E'\subset E$ and $e_0, e_1, ..., e_n\in E$,
\begin{itemize}
\item if $e_0\nearrow e_1\nearrow \cdots \nearrow e_n \nearrow e_0$ then $\{e_0, e_1, \cdots, e_n\}\in \#_a$,
\item if $E' \cup \{e_0\} \in \#_a$ and $e_0\le e_1$ then $E' \cup \{e_1\} \in \#_a$.
\end{itemize}
Informally, two events are in conflict if they cannot occur together in the same execution.

A LAES can be seen as a structure that encodes concisely several sequential executions; each of them being a linearization of the LAES.
\begin{definition}[Linearization]
  Let $\mathcal{E} = (E, L, \le, \nearrow, \Lambda)$ be a LAES.
  A \emph{finite linearization} of $\mathcal{E}$
  is a word $w = \Lambda(e_0)\ldots\Lambda(e_n)$ where the different $e_i\in E$ are distinct and such that:
  \begin{itemize}
  \item it is left-closed for causality: $$\forall e\in E, \forall e'\in \{e_0, \ldots, e_n\}, e\le e' \Rightarrow e\in \{e_0, \ldots, e_n\},$$
  \item the weak causality is respected: $$\forall i,j\in\{0,\ldots,n\},  e_i\nearrow e_j \Rightarrow i < j.$$
  \end{itemize}
  We denote $\lin{\mathcal{E}}$ as the set of all finite linearizations of $\mathcal{E}$.
\end{definition}

Let $(E, L, \le, \nearrow, \Lambda)$ be an 
asymmetric event structure and $e, e'\in E$ two events. We say that:
\begin{itemize}
\item $e$ is a \emph{cause} of $e'$, if $e$ happens before $e'$ in all executions;
\item $e$ is a \emph{weak cause} of $e'$, if there is no execution in which $e$ happens after $e'$;
\item $e$ and $e'$ are \emph{concurrent}, denoted $e || e'$, if they can occur in either order.
  Formally, $e||e'$ if neither $e\nearrow e'$ nor $e'\nearrow e$.
\item $e$ is \emph{preempted} by $e'$, denoted $e\rightsquigarrow e'$, if $e'$ can occur independently 
  from $e$, but after that, $e$ cannot occur anymore. Formally, $e\rightsquigarrow e'$ if 
  $e\nearrow e'$ and $e\not\le e'$.
\end{itemize}

\subsection{Rules}

\begin{figure}[t]
$$\inference[(\textsc{CauseYes})]
          {f\xrightarrow{k, l, c, a}_{i} f'}
          {\langle f, c', a'\rangle_L \xrightarrow{k, l, c\cup c', a \cup a' \cup c'}_{i} \langle f', c', a'\rangle_L}
          [$l\in L$]
$$
$$
\inference[(\textsc{CauseNo})]
          {f\xrightarrow{k, l, c, a}_{i} f'}
          {\langle f, c', a'\rangle_L \xrightarrow{k, l, c, a}_{i} \langle f', c', a'\rangle_L}
          [$l\not\in L$]
$$
\caption{The semantics of the new operator $\langle f, c, a\rangle_L$ is defined by two additional rules.}
\label{figure:rules_instr_cause}
\end{figure}

\begin{figure}[p]
\begin{center}
\begin{tabular}{lr}
\multicolumn{2}{l}{
\inference[(\textsc{Publish})]
          {\vwhite}
          {v\xrightarrow{k, !v, \emptyset, \emptyset}_{i} \langle \Stop, \{k\}, \emptyset\rangle_l}
          [\begin{tabular}{l}$v$ closed\\ $k$ fresh\end{tabular}]
}
\\
\inference[(\textsc{Stop})]
          {\vwhite}
          {\Stop \xrightarrow{k, \omega, \emptyset, \emptyset}_{i} \bot}
          [$k$ fresh]
&
\inference[(\textsc{StopCall})]
          {P\xrightarrow{k, \omega, c, a}_{i}\bot}
          {P(p)\xrightarrow{k, \omega, c, a}_{i}\bot}
\\
\multicolumn{2}{l}{
\inference[(\textsc{ExtStop})]
          {P\xrightarrow{k, !V, c, a}_{i} P' & p\xrightarrow{k', \omega, c', a'}_{i} p'}
          {P(p) \xrightarrow{k, \omega, c\cup c', a\cup a'}_{i} \bot}
}
\\
\multicolumn{2}{l}{
\inference[(\textsc{ExtCall})]
          {P\xrightarrow{k, !V, c, a}_{i} P' & p\xrightarrow{k', !v, c', a'}_{i} p'}
          {P(p) \xrightarrow{k, ?V_k(v), c\cup c', a\cup a'}_{i} \langle ?k, c\cup c' \cup \{k\}, a\cup a' \rangle_l}
}
\\
\multicolumn{2}{l}{
\inference[(\textsc{DefDeclare})]
          {[D/y]f \xrightarrow{k, l, c, a}_{i} f'}
          {D \# f \xrightarrow{k, l, c, a}_{i} f'}
          [$D$ is \textbf{def} $y(x) = g$]
}
\\
\multicolumn{2}{l}{
\inference[(\textsc{IntCall})]
          {P\xrightarrow{k, !D, c, a}_{i} P'}
          {P(p) \xrightarrow{k, ?D, c, a}_{i} \langle [D/y][p/x]g, c\cup \{k\}, a\rangle_l}
          [$D$ is \textbf{def} $y(x) = g$]
}
\\
\multicolumn{2}{r}{
\inference[(\textsc{ResT})]
          {?k \text{~receives~} T(v, c, a)}
          {?k\xrightarrow{j, !v, c, a\cup c}_{i} \langle \Stop, c\cup \{j\}, a \rangle_\omega}
          [$j$ fresh]
}
\\
\inference[(\textsc{ParLeft})]
          {f\xrightarrow{k, l, c, a}_{i} f'}
          {f|g\xrightarrow{k, l, c, a}_{i} f'|g}
          [$l\neq\omega$]
&
\inference[(\textsc{ResNt})]
          {?k \text{~receives~} NT(v, c, a)}
          {?k\xrightarrow{j, !v, c, a\cup c}_{i} ?k}
          [$j$ fresh]
\\
\inference[(\textsc{ParRight})]
          {g\xrightarrow{k, l, c, a}_{i} g'}
          {f|g\xrightarrow{k, l, c, a}_{i} f|g'}
          [$l\neq\omega$]
&
\inference[(\textsc{ResNeg})]
          {?k \text{~receives~} Neg(c, a)}
          {?k\xrightarrow{j, \omega, c, a\cup c}_{i} \bot}
          [$j$ fresh]
\\
\multicolumn{2}{l}{
\inference[(\textsc{ParStop})]
          {f\xrightarrow{k, \omega, c, a}_{i} f' & g\xrightarrow{k', \omega, c', a'}_{i} g'}
          {f|g \xrightarrow{k, \omega, c\cup c', a\cup a'}_{i} \bot}
}
\\
\multicolumn{2}{r}{
\inference[(\textsc{SeqV})]
          {f \xrightarrow{k, !v, c, a}_{i} f'}
          {f\seq{x}g \xrightarrow{k, h(!v), c, a}_{i} (f'\seq{x}g) | \langle [v/x]g, c\cup\{k\}, a \rangle_l}
}
\\
\inference[(\textsc{OtherV})]
          {f\xrightarrow{k, !v, c, a}_{i} f'}
          {f;g\xrightarrow{k, !v, c, a}_{i} f'}
&
\inference[(\textsc{SeqN})]
          {f\xrightarrow{k, n, c, a}_{i} f'}
          {f\seq{x}g \xrightarrow{k, n, c, a}_{i} f'\seq{x}g}
\\
\inference[(\textsc{OtherN})]
          {f\xrightarrow{k, n, c, a}_{i} f'}
          {f;g\xrightarrow{k, n, c, a}_{i} f';g}
&
\inference[(\textsc{SeqStop})]
          {f\xrightarrow{k, \omega, c, a}_{i} \bot}
          {f \seq{x} g \xrightarrow{k, \omega, c, a}_{i} \bot}
\\
\multicolumn{2}{l}{
\inference[(\textsc{OtherStop})]
          {f\xrightarrow{k, \omega, c, a}_{i} \bot}
          {f; g \xrightarrow{k, h(\omega), c, a}_{i} \langle g, c\cup\{k\}, a \rangle_l}
}
\\
\multicolumn{2}{l}{
\inference[(\textsc{PruneV})]
          {g \xrightarrow{k, !v, c, a}_{i} g'}
          {f\prune{x}g \xrightarrow{k, h(!v), c, a}_{i} \langle[\langle v, c\cup\{k\}, a \rangle_l/x]f, c\cup\{k\}, a \rangle_\omega}
}
\\
\multicolumn{2}{l}{
\inference[(\textsc{PruneN})]
          {g \xrightarrow{k, n, c, a}_{i} g'}
          {f\prune{x}g \xrightarrow{k, n, c, a}_{i} f\prune{x}\langle g', \emptyset, \{k\}\rangle_{!v}}
}
\\
\multicolumn{2}{l}{
\inference[(\textsc{PruneLeft})]
          {f\xrightarrow{k, l, c, a}_{i} f'}
          {f\prune{x}g \xrightarrow{k, l, c, a}_{i} f'\prune{x}g}
          [$l\neq\omega$]
}
\\
\multicolumn{2}{l}{
\inference[(\textsc{PruneStop})]
          {g \xrightarrow{k, \omega, c, a}_{i} \bot}
          {f \prune{x} g \xrightarrow{k, h(\omega), c, a}_{i} \langle[\langle \Stop, c\cup\{k\}, a \rangle_l/x]f, c\cup\{k\}, a \rangle_\omega}
}
\\
\end{tabular}
\end{center}
\caption{The instrumented version of the rules of the operational semantics.}
\label{figure:instrumented_semantics}
\end{figure}

Essentially, the instrumented semantics presented in Figure \ref{figure:instrumented_semantics} decorates the rules of
standard semantics, except that two rules are added (see Figure 
\ref{figure:rules_instr_cause}). The transition system defined by this instrumented semantics is
denoted $\rightarrow_i$ and the sequential executions starting from a program $f$ are
contained in the set $\llbracket f \rrbracket_i$.

Informally, the expression $\langle f, c, a\rangle_L$ 
evolves exactly like $f$, but some causes and weak causes may be added to the event.
For example, if $L=!v$
and $f$ produces an internal event, that is not a publication, only the rule 
\textsc{CauseNo} can be applied, so the instrumentation will have no effect. 
On the other hand, if $f$ publishes a value, the rule \textsc{CauseYes} applies
and $c$ and $a$ are added to the causes and weak causes of the publication. Note that
$c$ is also added to the weak causes. This is to ensure that causality is always a special 
case of weak causality.  

Let us now comment the most relevant instrumentations of the other rules.
Let us consider rule \textsc{SeqV}. When a value is published, a new instance of the 
right hand side expression is created. All the events produced by this new expression
need the former publication to \emph{have occurred before} them,
i.e. they are consequences of this publication. This is why the new
expression is instrumented.
Even if \textsc{PruneV} and \textsc{SeqV} are syntaxically very similar in their standard forms, 
the fact that both hand sides of the pruning operator are run in parallel makes them very different in terms of causality.
In the expression $(1|x)\prune{x} 2$, the second publication of $2$ is a consequence of the
first one, but not the publication of $1$. This is why the instrumentation covers the occurrences
of the newly bound variable. However, this is not sufficient. Consider the program
$(\Stop\prune{x}2); 3$. The publication of $3$ must wait the end of
the left hand side (i.e the publication of $2$). However, this
publication is
useless, in the sense that no variable $x$ can be bound to its value. To handle this case, we add an 
instrumentation to the whole expression that is only triggered when
the expression stops. 
Finally, the rule \textsc{PruneN} is also interesting as it generates weak causality. Indeed,
in the program $x\prune{s}((1+1)|3)$, the left hand side can call site $+$ and then publish $3$,
or publish $3$ directly, but can never publish $3$ and then call site $+$, because a publication 
preempts any other event. Of course, it could also wait for the answer of the site and then publish
$2$, which would preempt the publication of $3$. This preemption relation is operated by an
instrumentation that contains $k$ as weak causes and that is triggered only in case of publication.

\subsection{Concurrent executions}

The equivalent of traces in the instrumented semantics are the
{\em concurrent executions}, represented by LAES.

\begin{definition}[Concurrent execution]
  Let $\sigma = \sigma^1\ldots\sigma^n \in \llbracket f_0 \rrbracket_i$, where for all $i$, $\sigma^i = (\sigma_k^i, \sigma_l^i, \sigma_c^i, \sigma_a^i)$.
  We define the \emph{concurrent execution} of $\sigma$ as the LAES:
  $$\bibar{\sigma} = (\{\sigma_k^1,\ldots, \sigma_k^n\}, \{\sigma_l^1,\ldots, \sigma_l^n\}, \le, \nearrow, \Lambda)$$
  where for all $i, j\in \{1,\ldots, n\}$:
  \begin{itemize}
  \item $\sigma_k^i \le \sigma_k^j$ if $\sigma_k^i \in \sigma_c^j$ or $i=j$,
  \item $\sigma_k^i \nearrow \sigma_k^j$ if $\sigma_k^i \in \sigma_a^j$,
  \item $\Lambda(\sigma_k^i) = \sigma_l^i$.
  \end{itemize}
\end{definition}
As the fields $\sigma_c^i$ and $\sigma_a^i$ only contain events that happened before $\sigma^i$ in the sequential execution, 
both $\le$ and $\nearrow$ are order relations and every event has a finite causal history. For the same reason, 
$\nearrow$ is acyclic. Moreover, it is easy to check that weak causality is more general than causality, so this definition 
actually corresponds to a real LAES. 

We now state the main result: the behavior of a program is preserved by the instrumented 
semantics. It is established through two properties. The first one justifies the name of the 
instrumented semantics and the second one proves that the instrumentation is correct, 
i.e. that it does not define incorrect behaviors. Note that we do not give a complete proof
of the two propositions for lack of space, but it can be found in \cite{rr}. 

\begin{proposition}[Instrumentation]\label{prop:instrumentation}
  The projections of the executions produced by the instrumented semantics on their labels correspond exactly to 
  the executions of the standard semantics:
  $$\forall f \in Orc, \{\sigma_l^1... \sigma_l^n | \sigma\in \llbracket f \rrbracket_i\} = \llbracket f \rrbracket.$$
  In other words, it is always possible
  to instrument a standard execution to get a concurrent execution, and conversely
  we can get a standard execution from an instrumented execution by a simple projection.
\end{proposition}
The proof of this property is straightforward since both semantics contain similar rules. 
The only difficulty comes from the applications of \textsc{CauseYes} and \textsc{CauseNo}, 
that slightly modifie the structure of the derivation trees. All in all, both executions 
are similar.

\begin{proposition}[Correctness]\label{prop:instrument_correct}
  The linearizations that can be inferred from an execution in the instrumented semantics are correct
  with respect to the standard semantics:
  $$\forall f \in Orc, \forall \sigma\in \llbracket f \rrbracket_i, \lin{\bibar{\sigma}}\subset \llbracket f \rrbracket.$$
\end{proposition}
This proof is much more complicated. Let us consider a linearization $L \in \lin{\bibar{\sigma}}$. 
To prove that $L\in \llbracket f \rrbracket$, we show that it is possible to progressively transform 
$\sigma_l^1... \sigma_l^n$ into $L$ by applying a series of small steps that correspond either to the 
inversion of two consecutive concurrent events, to the preemption of an event by its successor or to
a prefixation of the sequence. As $\sigma_l^1... \sigma_l^n \in \llbracket f \rrbracket$ and each step 
preserves this property, we get that $L \in \llbracket f \rrbracket$. The main difficulty concerns the 
proof of the correctness of the two first steps, as it requires a proof for all pairs of possible 
consecutive rules.

By introducing concurrency and preemption between events that were 
arbitrarily ordered by the standard semantics, the instrumented semantics 
gathers many sequential executions into one concurrent execution, which 
hugely reduces the number of different executions. However, all the events
contained in a concurrent execution are also contained into a single 
sequential execution. Therefore, no instrumentation is able to capture
conflict, as two conflictual events would never occur together. 
This is why completeness cannot be achieved with this approach.

\section{Application}
\label{section:application}

\begin{figure}[t]
\newcommand{\lineindent}{\\\phantom{'}\hspace{\fill}}
  \begin{multicols}{2}
    (1, $l_{1}$, $\emptyset$, $\emptyset$)\\
    (2, $l_{2}$, $\emptyset$, $\emptyset$)\\
    (3, $l_{3}$, $\emptyset$, $\emptyset$)\\
    (4, $l_{4}$, $\emptyset$, $\emptyset$)\\
    (5, $l_{5}$, \{1\}, \{1\})\\
    (6, $l_{6}$, \{4\}, \{4\})\\
    (7, $l_{7}$, \{1,4-6\}, \{1,4-6\})\\
    (8, $l_{8}$, $\emptyset$, $\emptyset$)\\
    (9, $l_{9}$, \{1,5,8\}, \{1,5,8\})\\
    (10, $l_{10}$, \{1\}, \{1\}) \\
    (11, $l_{11}$, \{1,8,10\}, \{1,8,10\}) \\
    (12, $l_{12}$, \{1,4-7\}, \{1,4-7\}) \\
    (13, $l_{13}$, \{1,4,6,10\}, \{1,4,6,10\}) \\
    (14, $l_{14}$, \{1,4,6,10,13\}, \{1,4,6,10,13\}) \\
    (15, $l_{15}$, \{1,4-7,12\}, \{1,4-7,12\}) \\
    (16, $l_{16}$, \{1,4,6,10,13,14\}, \{1,4,6,10,13,14\}) \\
    (17, $l_{17}$, \{1,4,6,10,13,14,16\}, \lineindent\{1,4,6,10,13,14,16\}) \\
    (18, $l_{18}$, \{1,4-7,12,15\}, \{1,4-7,12,15\}) \\
    (19, $l_{19}$, \{1,3-7,12,15\}, \{1,3-7,12,15\}) \\
    (20, $l_{20}$, \{1,3,4,6,10,13,14,16\}, \lineindent\{1,3,4,6,10,13,14,16\})

    (21, $l_{21}$, \{2\}, \{2, 1\}) \\
    (22, $l_{22}$, \{1-7,10,12-16,19-21\}, \lineindent\{1-7,10,12-16,19-21\}) \\
    (23, $l_{23}$, \{1-7,10,12-19,22\}, \lineindent\{1-7,10,12-19,22\}) \\
    (24, $l_{24}$, \{1-7,10,12-19,22,23\}, \lineindent\{1-7,10,12-19,22,23\}) \\
    (25, $l_{25}$, \{1-7,10,12-19,22,23\}, \lineindent\{1-7,10,12-19,22,23\}) \\
    (26, $l_{26}$, \{1-7,10,12-19,22-24\}, \lineindent\{1-7,10,12-19,22-24\}) \\
    (27, $l_{27}$, \{2\}, \{2, 1\}) \\
    (28, $l_{28}$, \{1,2,5,9-11,27\}, \{1,2,5,9-11,27\}) \\
    (29, $l_{29}$, \{1,2,5,9-11,27\}, \{1,2,5,9-11,27\}) \\
    (30, $l_{30}$, \{1,2,4-7,9-18,27-29\}, \lineindent\{1,2,4-7,9-18,27-29\}) \\
    (31, $l_{31}$, \{1,2,4-7,9-18,27-30\}, \lineindent\{1,2,4-7,9-18,27-30\}) \\
    (32, $l_{32}$, \{1,2,4-7,9-18,27-30\}, \lineindent\{1,2,4-7,9-18,27-30\}) \\
    (33, $l_{33}$, \{1,2,4-7,9-18,27-31\}, \lineindent\{1,2,4-7,9-18,27-31\}) \\
  \end{multicols}
\caption{An execution augmented with causal information. Numbers refer to
the events of Figure~\ref{figure:sequential_execution}.}
\label{figure:instrumented_execution}
\end{figure}

Let us reuse the example presented in Section \ref{section:orc} and
the execution of Figure \ref{figure:sequential_execution}. Figure
\ref{figure:instrumented_execution} shows the trace augmented with the
causal information gained by the instrumented semantics. 

\begin{figure}[t]

\centering
\includegraphics[width=.8\textwidth]{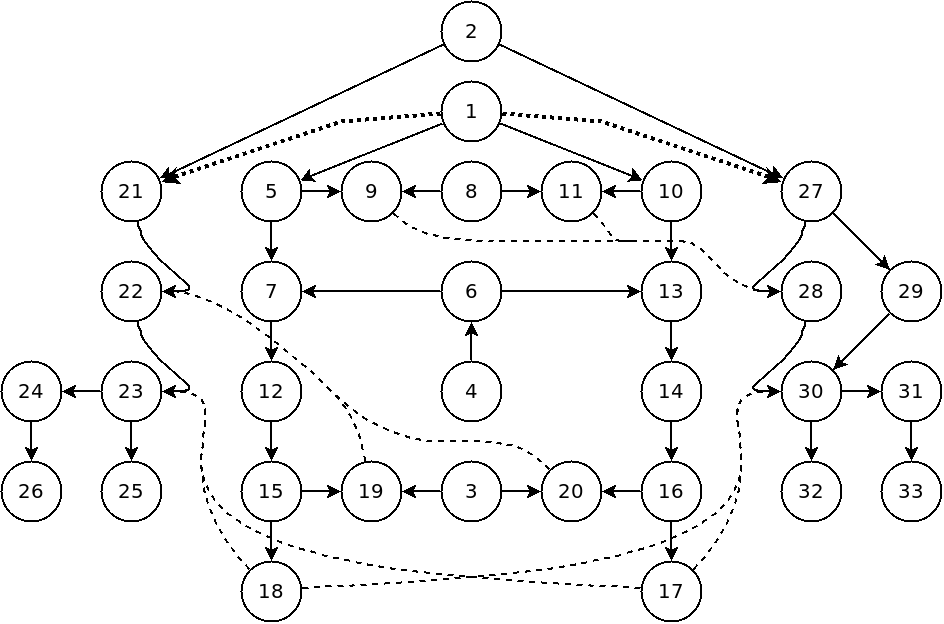}
\caption{Corresponding LAES in a graphical form.}
\label{figure:graphical_instrumented_execution}
\end{figure}

Figure \ref{figure:graphical_instrumented_execution} shows the
LAES in its graphical form. Events taht correspond to site calls are
represented in circles and connected by three kinds of arrows.
Direct causality is figured by solid and dashed arrows. Solid arrows
represent
program causality, that is specified by the instrumented semantics, while
dashed
arrows represent data causes, that are managed by the sites themselves. A
\emph{call}
to a write on a site is a cause of the \emph{publications} of the next
read on this site,
so the write is a cause of all the consequences of the read.
Moreover, preemption is figured by dotted arrows. The call to \verb+each()+
--- and all its consequences --- is preempted by the publication made by
\verb+timer(2000)+ --- and all its consequences.

\paragraph{Root causes analysis.} This execution concurrently raises 
two alarms. Let us consider their last common causes, i.e. the events that are 
causes of both alarms, and that are not causes of another such event. 
The alarms have two last common causes: \verb+timer(2000)+ and \verb+best_offer.write(O1)+.
The timer is not to blame here, as it has no causes and is just used as a starter
for the program. Indeed, \verb+best_offer.write(O1)+ is the root cause for these two alarms.
If this event did not exist, the value published by \verb+best_offer.read()+ would be \verb+O2+
for both calls, \verb+A2.exists(O2)+ and \verb+=(02, O2)+ would be true and no alarm would be raised.

\paragraph{Detection of Race Conditions.} We can see that the events \verb+best_offer.write(O1)+
and \verb+best_offer.write(O2)+ are concurrent, as well as \verb+best_agency.write(O1)+
and \verb+best_agency.write(O2)+. In this context, these events can interleave so that
\verb+best_offer+ and \verb+best_agency+ get inconsistent values.

\section{Conclusion}\label{section:conclusion}

We based our work on the Orc core calculus, 
as it is expressive enough to easily generate many situations found in distributed systems, 
such as causality, concurrency and preemption, and remains simple enough to be tractable
in a formal work. Our contribution consists of an instrumentation of the standard structural 
operational semantics of Orc that tracks causality and weak causality at 
runtime to build LAES, well suited to represent concurrent executions.
We think LAES are an interesting tool to access important properties of 
orchestrations. We illustrate this point on two questions: 
root cause analysis and detection of race conditions.
Beyond the Orc language, we think that the article presents a general approach
that can be used for other non-deterministic languages with concurrency operators.
Based on this work, we think it is possible to produce the same
information using a source to source transformation.
Such a technique would be easier to implement, as it does not require to modify
the execution engine of the language.

\bibliographystyle{splncs}

\end{document}